\documentclass[letter]{aa} 
\usepackage{graphicx}
\usepackage{txfonts}
\usepackage{natbib}
\usepackage{float}
\usepackage[flushleft]{threeparttable}
\bibpunct{(}{)}{;}{a}{}{,} 
\usepackage[normalem]{ulem}
\newcommand{\modif}[1]{\textnormal{#1}}
\newcommand{\modiff}[1]{\textnormal{#1}}

\newcommand{\modiflet}[1]{\textnormal{#1}}

\begin{document}

   \title{Multi-wavelength VLTI study of the puffed-up inner rim of a circumbinary disc \thanks{Based on observations collected at the European Southern Observatory under ESO programmes 094.D-0865, 0102.D-0760, 60.A-9275, and 0104.D-0739}}

   \author{A. Corporaal \inst{1}
           \and  J. Kluska \inst{1}
           \and H. Van Winckel \inst{1}
           \and D. Bollen \inst{1,2,3}
           \and D. Kamath \inst{2,3}
           \and M. Min \inst{4}}

   \institute{Institute of Astronomy, KU Leuven,
              Celestijnenlaan 200D, 3001 Leuven, Belgium\\
             \email{akke.corporaal@kuleuven.be}
    \and Department of Physics \& Astronomy, Macquarie University, Sydney, NSW 2109, Australia
    \and Astronomy, Astrophysics and Astrophotonics Research Centre, Macquarie University, Sydney, NSW 2109, Australia
    \and SRON Netherlands Institute for Space Research, Sorbonnelaan 2, 3584 CA, Utrecht, The Netherlands}

   \date{Received ; accepted}

 
  \abstract
   {The presence of stable, compact circumbinary discs of gas and dust around post-asymptotic giant branch (post-AGB) binary systems has been well established. We focus on one such system, IRAS\,08544-4431.}
   {We present an interferometric multi-wavelength analysis of the circumstellar environment of IRAS\, 08544-4431. The aim is to constrain different contributions to the total flux in the H-, K-, L-, and N-bands in the radial direction.}
   {The data \modiflet{obtained with the three current instruments on the Very Large Telescope Interferometer (VLTI), VLTI/PIONIER, VLTI/GRAVITY, and VLTI/MATISSE,} range from the near-infrared, where the post-AGB star dominates, to the mid-infrared, where the disc dominates. We fitted the following two geometric models to the visibility data to reproduce the circumbinary disc: a ring with a Gaussian width and a flat disc model with a temperature gradient. The flux contributions from the disc, the primary star (modelled as a point source), and an over-resolved component were recovered along with the radial size of the emission, the temperature of the disc as a function of radius, and the spectral dependencies of the different components.}
   {The trends of all visibility data were well reproduced with the geometric models. The near-infrared data were best fitted with a Gaussian ring model, while the mid-infrared data favoured a temperature gradient model. This implies that \modif{a vertical structure is present at the disc inner rim, which we attribute to a rounded puffed-up inner rim}. The N-to-K size ratio is \modif{2.8}, referring to a continuous flat source, analogues to young stellar objects.}
   {By combining optical interferometric instruments operating at different wavelengths, we can resolve the complex structure of circumstellar discs and study the wavelength-dependent opacity profile. A detailed radial, vertical, and azimuthal structural analysis awaits a radiative transfer treatment in 3D to capture all non-radial complexity.}


   \maketitle
%

\section{Introduction}
Circumstellar discs have been found at different evolutionary stages of stars. Here we focus on circumbinary discs around evolved post-asymptotic giant branch (post-AGB) binary systems. The presence of these discs has been well established \citep[e.g.][]{vanwinckel_2003, vanWinckel_2017}. Observational evidence of such discs is a flux excess starting at near-infrared (near-IR) wavelengths seen in the spectral energy distributions (SEDs) of these targets, indicating that circumstellar dust must extend from the dust sublimation radius outwards.

The velocity field of some objects has been spatially resolved at millimetre wavelengths in CO, using the Atacama Large Millimeter/submillimeter Array (ALMA) and Plateau de Bure Interferometers \citep{Bujarrabal_2013, Bujarrabal_2015, Bujarrabal_2017, Bujarrabal_2018}, showing that the discs are in Keplerian rotation. Other indicators of longevity come from observations of strong dust grain processing in the form of a high degree of crystallinity \citep{Gielen_2011}, the presence of large grains \citep[e.g.][]{Gielen_2011}, and single-dish CO-line observations that confirm that rotation is widespread \citep{Bujarrabal_2013a}. 

The apparent size of the disc is compact. Optical interferometric techniques are, thus, needed to probe the compact infrared emission from the inner regions of such discs. In this contribution we present our interferometric multi-wavelength approach for one specific post-AGB binary system: IRAS\,08544-4431 (hereafter IRAS\,08544), located at an approximate distance of $1.53 \pm 0.12$ kpc \citep{Bailer-Jones_2021}. This system has been studied in \citet{Hillen_2016}, in which a successful interferometric imaging experiment is presented with data obtained with PIONIER ($H$-band) and in \citet{Kluska_2018}, in which a radiative transfer approach was used to explain the inner rim emission in the PIONIER data. These studies show that the inner rim of the disc \modiflet{is} well resolved. The components that contribute to the $H$-band flux are the central star, the accretion disc around the companion, the circumbinary disc, and an over-resolved component of unknown origin.

While in the $H$-band, the hot, optically thick disc inner rim is probed, the mid-IR emission comes from a more extended region. We investigate the contributions of the different components at different wavelengths by using a combination of interferometric data in four bands taken with all the current instruments on the \modiflet{Very Large Telescope Interferometer (VLTI)}. We focus on analysing the visibility data of IRAS\,08544 by combining, for the first time, data from PIONIER, GRAVITY, and MATISSE to study the circumbinary disc in the radial direction with geometric models. The goal is to physically understand the flux contributions from the different components from the near-IR to the mid-IR where the SED shifts from a star-dominated to a disc-dominated regime.

In this letter, we present our multi-wavelength interferometric study of a disc around an evolved system using all current instruments on the VLTI. In Sect. \ref{Sec2} we present the data. In Sect. \ref{Sec3} we discuss the geometric models used for interpreting these data and we analyse the results in Sect. \ref{Sec4}. We discuss our findings in Sect. \ref{Sec5} and conclude in Sect. \ref{Sec6}. 

\section{Observations}
\label{Sec2}
Interferometric observations were obtained with the following three current four-telescope beam combiner instruments on ESO's VLTI at Mount Paranal in Chile: PIONIER \citep{LeBouquin2011}, GRAVITY \citep{GRAVITY2017}, and MATISSE \citep{Lopez2014}. The instruments provide six simultaneous baselines and three independent closure phases per measurement. In the following we describe the obtained data per instrument.

\subsection{PIONIER}
PIONIER is operating in the $H$-band (wavelengths between 1.5 $\mu$m and 1.85 $\mu$m). The data were taken on the nights of 2015 January 21, 2015 January 24, and 2015 February 23 (prog. ID: 094.D-0865, PI: Hillen), using the three configurations on the 1.8m Auxiliary Telescopes (ATs). The data set consists of 828 squared visibility data points. For a log of the observations, readers can refer to \citet{Hillen_2016}.

\subsection{GRAVITY}
GRAVITY is operating in the $K$-band (wavelengths between 2.0 and 2.4 $\mu$m). Observations were taken on the nights of 2018 November 26, 2018 December 23, and 2019 January 15 (prog. ID: 0102.D-0760, PI: Bollen), with the three configurations of the ATs at high resolution (R$\sim$ 4000) in single field mode. The data were reduced and calibrated with the GRAVITY pipeline version 1.1.2, resulting in 83372 squared visibility data points. Table \ref{GRAVITY_obs} summarises the log of the observations. 

\subsection{MATISSE}
MATISSE has been available to the community since 2019, starting with a science verification (SV) phase. MATISSE operates in the $L$ (2.9 - 4.2 $\mu$m), $M$ (4.5 - 5.0 $\mu$m), and $N$-bands (8 - 13 $\mu$m). During the SV phase and ESO's P104, $M$-band observations were not yet offered. The data sets of MATISSE consists of data taken with the three configurations of the ATs on the nights of 2019 April 29, 2019 May 01, 2019 May 02, 2019 May 04, 2019 May 09, and 2019 May 11 (prog. ID 60.A-9275, PI: Kluska) during the SV phase of MATISSE and on 2020 January 21 and 2020 March 01 (prog. ID 0104.D-0739, PI: Kluska). The log of the observations are displayed in Table \ref{MATISSE_obs}. All observations were performed in hybrid mode at low resolution.

The data were reduced and calibrated with the MATISSE data reduction pipeline version 1.5.5 and 1.5.8 for the SV phase data and the P104 data, respectively. We only took data into account for which the data reduction and calibration could be done successfully and for which no observational problems occurred.

The data on the wavelength edges of both bands could not be used in further analysis since these data showed unstable behaviour. The final data set consists of 3432 and 3762 visibility data points for the $L$- and $N$-band, respectively. The data of the $N$-band were reduced in correlated flux mode of the MATISSE data reduction pipeline. We note that $N$-band photometry was not taken for all nights during the SV, meaning that the coherent flux measurements could not be normalised and therefore visibilities could not be determined. The reduced and partially calibrated correlated flux data \modif{from the Beam Commuting Device were merged} and subsequently multiplied by the theoretical blackbody intensity of the calibrator (see Table \ref{N-band_cal_info}).
\\
\\
Fig. \ref{uv_coverages} shows an overview of the visibility data and the uv-coverages of the accepted data of each band. The visibility data show a strong chromatic effect in the near-IR, while this is not seen at mid-IR wavelengths, where the stellar contribution has significantly decreased. The mid-IR data \modif{show} bumps, reminiscent of Bessel functions.

\section{Geometric models}
\label{Sec3}

\begin{table*}[t]
\caption{Best-fit parameters of the geometric models}
\label{best_fits_geometric_models}
\begin{threeparttable}
\begin{tabular}{p{1.21 cm}ccccccccc}
\hline \hline
& \multicolumn{2}{c}{$H$-band (1.65 $\mu$m)}&  \multicolumn{2}{c}{$K$-band (2.15 $\mu$m)} & \multicolumn{2}{c}{$L$-band (3.5 $\mu$m)} &
\multicolumn{2}{c}{$N$-band (10.5 $\mu$m)}\\ 

& Ring & Flat disc & Ring &Flat disc & Ring & Flat disc & Ring& Flat disc \\
\hline
Parameter\\
\hline

f$_{star_0} (\%)$ & $62.3 \pm 0.1$& $62.4 \pm 0.1$ &$31.7 \pm 0.4$ & $32.1 \pm 0.5 $ &$ 4.4 \pm 0.1$& $6.1 \pm 0.1$&$3.3 \pm 1.0$ & $2.9 \pm 0.4$\\
f$_{back_0} (\%)$ & $13.6 \pm 0.3$ & $13.4 \pm 1.3$&$20.6 \pm 0.5$& $10.3 \pm 0.6 $&$21.0 \pm 0.1$& $10.0 \pm 0.3$&-& -\\
$d_\mathrm{bg}$ & $1.06\pm 0.15$ & $1.11 \pm 0.40$&$ 0.45 \pm 0.13$ &$1.71 \pm 0.60$ &$0.002 \pm 0.891 $& $3.62 \pm 0.25$ &-& -\\
$\theta$ (mas) & $14.23 \pm 0.04$ & -& $14.56\pm 0.09 $ & - & $15.85 \pm 0.04$ & -& $23.7\pm 2.2 $ & -\\
$w$ (mas) & $3.63 \pm 0.11$ & -& $4.54 \pm 0.07$ &- &$6.88 \pm 0.12$ & -&$22.4 \pm 4.0$& -\\
$T_\mathrm{ring}$ (K) & $1286 \pm 73$ & -& 1035 $\pm 61$& - & $839 \pm 109$&- &$636 \pm 289$&-\\
$T_\mathrm{in}$ (K)& -& $1498 \pm 32$ & -&$1426 \pm 116$ & - &$1504 \pm 89$ & - &$1550 \pm 198$\\
$q$ & -& $0.72 \pm 0.07$& -&$0.60 \pm 0.05$& -& $0.71 \pm 0.04$ &-& $0.80 \pm 0.07$ \\
$R_\mathrm{in} \mathrm{(mas)}$& - &$5.99 \pm 0.09$ &- &$5.45 \pm 0.02$& -& $5.47 \pm 0.03$ &-& $6.53 \pm 0.20$\\
$F_\mathrm{tot}$ (Jy) & - & - & - & - & - &- & $165 \pm 8$& $210 \pm 20$\\
$\chi^2_\mathrm{red}$ & 4.67& 5.59 & 24.9 &29.7 &65.2 & \modif{25.4} &2.65&\modif{1.43}\\
\hline
\end{tabular}

\end{threeparttable}
\end{table*}

In order to fit the visibility data with geometric models, the wavelength-dependent integrated flux contribution of each component needs to be recovered. The considered components are a single star, an inclined disc, and an over-resolved emission. To keep our models simple, we did not include the contribution of the accretion disc around the secondary in the modelling, despite the target being a confirmed binary \citep{Maas_2003}, meaning that the central luminous source is not in the centre, but it revolves around the centre of mass. The contribution of the accretion disc around the secondary is only 3.9\% in the $H$-band \citep{Hillen_2016, Kluska_2018}. Another simplification is that the disc geometry is assumed to be azimuthally symmetric, while significant azimuthal variations were found in the PIONIER data by \citet{Hillen_2016} and \citet{Kluska_2018}.

The flux fractions of the three components, $f_0^i$, are defined at the central wavelength $\lambda_0$ of each band such that $\sum f_0^i$($\lambda = \lambda_0)=1$. The flux of the primary is described by the photospheric flux of the SED of IRAS\,08544 \modif{\citep{Kluska_2018}} normalised to unity at the central wavelengths of the bands. The over-resolved component is the flux outside of the modelled disc and the stellar flux. This background flux is modelled by the following:
\begin{equation}
f_\mathrm{back}=f_\mathrm{back_0}\left(\frac{\lambda}{\lambda_0}\right)^{-d_\mathrm{bg}}
,\end{equation}
where $d_\mathrm{bg}=\frac{d \log F_\lambda}{d \log \lambda}$ is the index of the background emission.

The circumbinary disc was modelled using the following two geometric models: Gaussian ring models and optically thick flat disc models, as mathematically described by \citet{Kluska_2019} and \citet{Menu_2015}, respectively. These models were successfully applied to modelling interferometric data of circumbinary discs around post-AGB binaries in the near-IR by, for example, \citet{Hillen_2016}, \citet{ Kluska_2019}, and by \citet{Hillen_2017} in the mid-IR. We applied both geometric models in an attempt to fit the visibility data for all bands. The \modif{position angle of the major axis of the disc, measured north to east, and the} inclination were not fitted to keep the model simple. Those values were taken from the best fits of \citet{Hillen_2016} and assumed to be $6^\circ$ and $19^\circ$, respectively.

The Gaussian ring model establishes the size of the emission, $\theta$, which has a certain Gaussian width, $w$, and a single ring temperature, $T_\mathrm{ring}$. In the flat disc model, the disc is divided into several rings. There is an inner cavity, a sharp inner rim at a radius $R_\mathrm{in}$, and a brightness distribution determined by the temperature structure parameterised with the radius. The disc surface emits blackbody radiation such that the intensity distribution is given by
\begin{equation}
    I_\nu\propto B_\nu\Big(T_\mathrm{in}\Big(\frac{r}{R_\mathrm{in}}\Big)^{-q}\Big)  \quad \quad \mathrm{for}  \quad R_\mathrm{in}<r<R_\mathrm{out}
    \label{Eq: I_nu}
\end{equation}
and zero elsewhere. \modiflet{Here} $T_\mathrm{in}$ is the temperature of the inner rim, $R_\mathrm{out}$ is the outer radius of the disc, and $q$ is the exponent of the temperature profile. \modiflet{We set $R_\mathrm{out}$} to 130 mas (200 au), but the result is independent of the exact value.

In classical, flaring, protoplanetary disc models, the temperature profile is $T\propto r^{-1/2}$, while the steepest profile is expected for passively irradiated discs in the geometrically thin limit and corresponds to $T\propto r^{-3/4}$ \citep{KenyonHartmann_1987}. The value of $q$, however, also depends on the optical depth, which is assumed to be constant here; $q$ cannot be directly attributed to the true temperature gradient in the disc, but it acts to distribute the flux over radial annuli \modif{\citep[see e.g.][] {Menu_2015}}. 

The visibility drop at short baselines determines the inner rim radius and therefore the size, and it is important that the models reproduce it. We, therefore, required the model to fit the short baselines in all bands. The shortest piece of baseline data is underrepresented in the $H$-, $K$-, and $L$-bands. In order to fit those data points, more weight was given to some short baseline data points by increasing the weight of those points by a factor of 9 for the $H$-band ($B< 8$ M$\lambda$) and $K$-band ($B< 5.5$ M$\lambda$), and by a factor of 25 for the $L$-band ($B< 3.5$ M$\lambda$) in both models. The effect of this is shown in Fig. \ref{short_baselines_comp} for the Gaussian ring models. The $N$-band data had the same problem \modif{when fitting the short baseline data} in the Gaussian ring model such that the weight of the shortest baselines ($B< 2.5$ M$\lambda$) was increased by a factor of 9, while the flat disc model could fit the shortest baselines without such a modification. 

The data were first fitted using a Levenberg-Marquardt fitting routine (see Table \ref{ranges of parameters} for the accepted ranges of the fitting parameters). The best-fit values were subsequently used as a starting point for a Markov Chain Monte Carlo (MCMC) minimisation routine to determine the uncertainties on the parameters\modif{, except for the values of $T_\mathrm{in}$ for the $L$- and $N$-bands, for which we assumed a starting point of 1500 K}. The reported values are the MCMC results from the Python package \textsc{lmfit} and are given with respect to the central wavelength of the band. For the $H$-, $K$-, and $L$-band, six free parameters were fitted for the Gaussian ring model: $\theta$, $w$, and $T_\mathrm{ring}$ for the disc along with $f_\mathrm{star_0}$, $f_\mathrm{back_0}$, and $d_\mathrm{bg}$ for the other flux contributions. For the flat disc model, the latter three parameters were fitted along with $T_{\mathrm{sub}}$, $q$, and $R_\mathrm{in}$ for the disc. Since the $N$-band data were reduced in correlated flux mode, the total flux at zeroth baseline, $F_\mathrm{tot}$, was also fitted for these data instead of $f_\mathrm{back_0}$ and $d_\mathrm{bg}$ as these data \modiff{are} not sensitive to the over-resolved emission. Since both models had the same number of degrees of freedom, the best-fitting model of each band was determined from the reduced $\chi^2$. This quantity was also used to assess the quality of the fits.

\begin{figure*}
\centering
  \includegraphics[width=17cm]{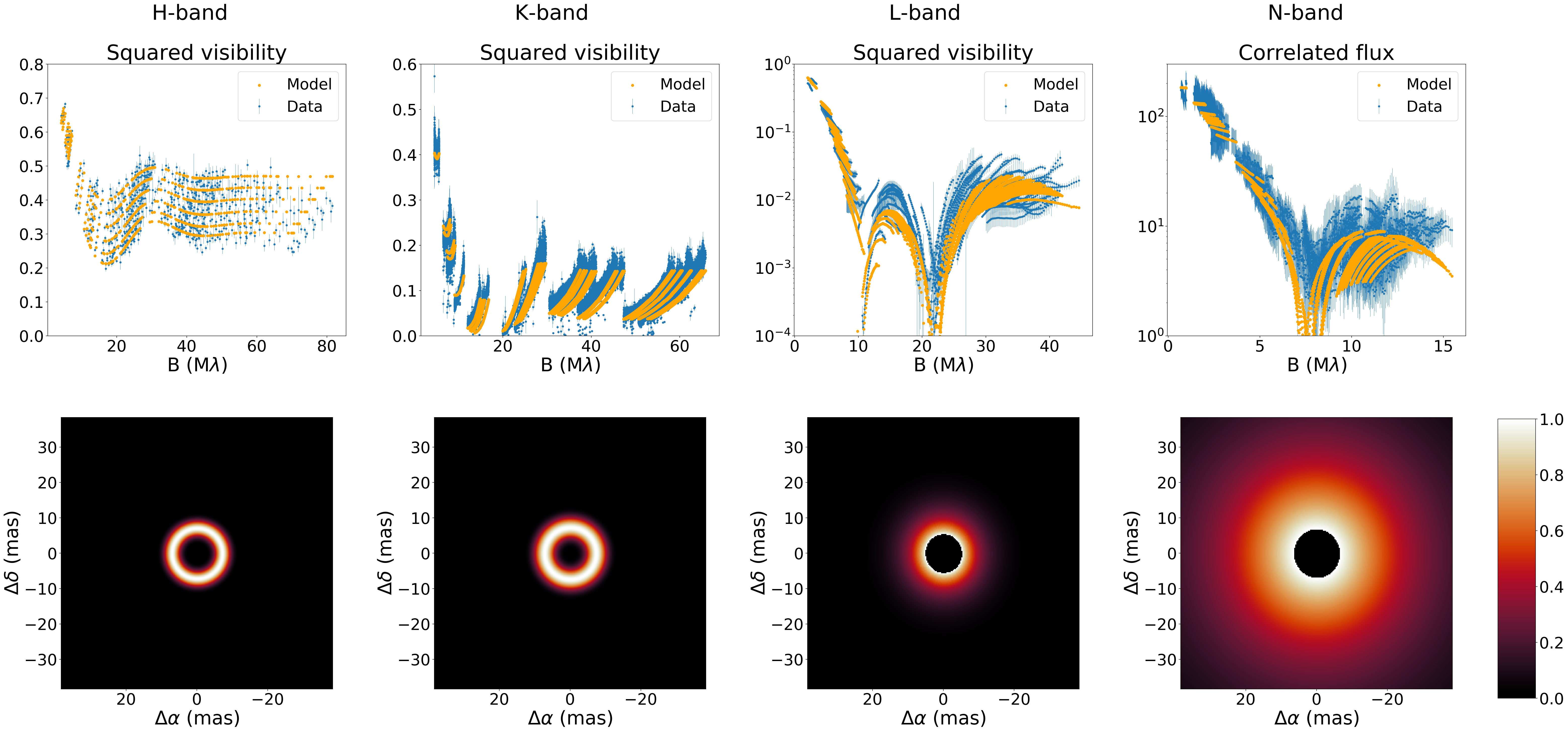}
  
  \caption{\textit{Top}. Visibility data of (from left to right) the $H$-, $K$-, $L$-, and $N$-band in blue and the best fitting geometric model in orange. The correlated flux measurements in the $N$-band are given in units of Jansky. \textit{Bottom}. Images of the circumbinary disc from the best fitting geometric model for each band normalised to the total flux of each individual image.}
  \label{fig:Images_all_bands}
\end{figure*}

\section{Results}
\label{Sec4}
The best-fit values of the two geometric models that reproduce the visibility data of each band are listed in Table \ref{best_fits_geometric_models}. Fig. \ref{fig:Images_all_bands} shows the visibility data compared to the corresponding best-fit synthetic visibilities and the corresponding model images of the disc. \modif{Fig. \ref{cut_center_Fig1} shows the radial cuts of these best-fit model images.} From the $\chi^2_\mathrm{red}$, it can be perceived that the best-fitting model to the visibility data is the Gaussian ring model for the $H$- and $K$-band and the flat disc model for the $L$- and $N$-band.

The stellar contribution is well-constrained as both geometric models predict similar contributions. The stellar contribution to the total flux budget decreases from $62.3\%$ in the near-IR to $2.9\%$ in the mid-IR.

The amount of \modif{background flux} needed to reproduce the visibilities \modif{strongly depended}  on the model and varied between 10-20\%. We note that the different bands probe a different field-of-view such that the amount of \modif{background} flux captured may vary from band to band: $\sim 250$ mas in $H$, $\sim 330$ mas in $K$, $\sim 530$ mas in $L$, and $\sim 1600$ mas in $N$.

It is important to note that $d_\mathrm{bg}$ could not be constrained for the Gaussian ring model in $L$ \modiff{(as revealed by the large uncertainty on this parameter in Table 1)}, since the V$^2$ predicted by the model are too low in the longer baselines ($B>25$ M$\lambda$) (see Fig. \ref{short_baselines_comp}). \modif{The Gaussian ring} model predicts a higher \modif{background} flux ratio \modif{compared to the flat disc model} because of the geometry of the model. The contribution of the \modif{background} flux in the flat disc model is found to be rather constant over the wavelength range of the $H$-, $K$-, and $L$-bands. 

\begin{figure}[b]
  \resizebox{\hsize}{!}{\includegraphics{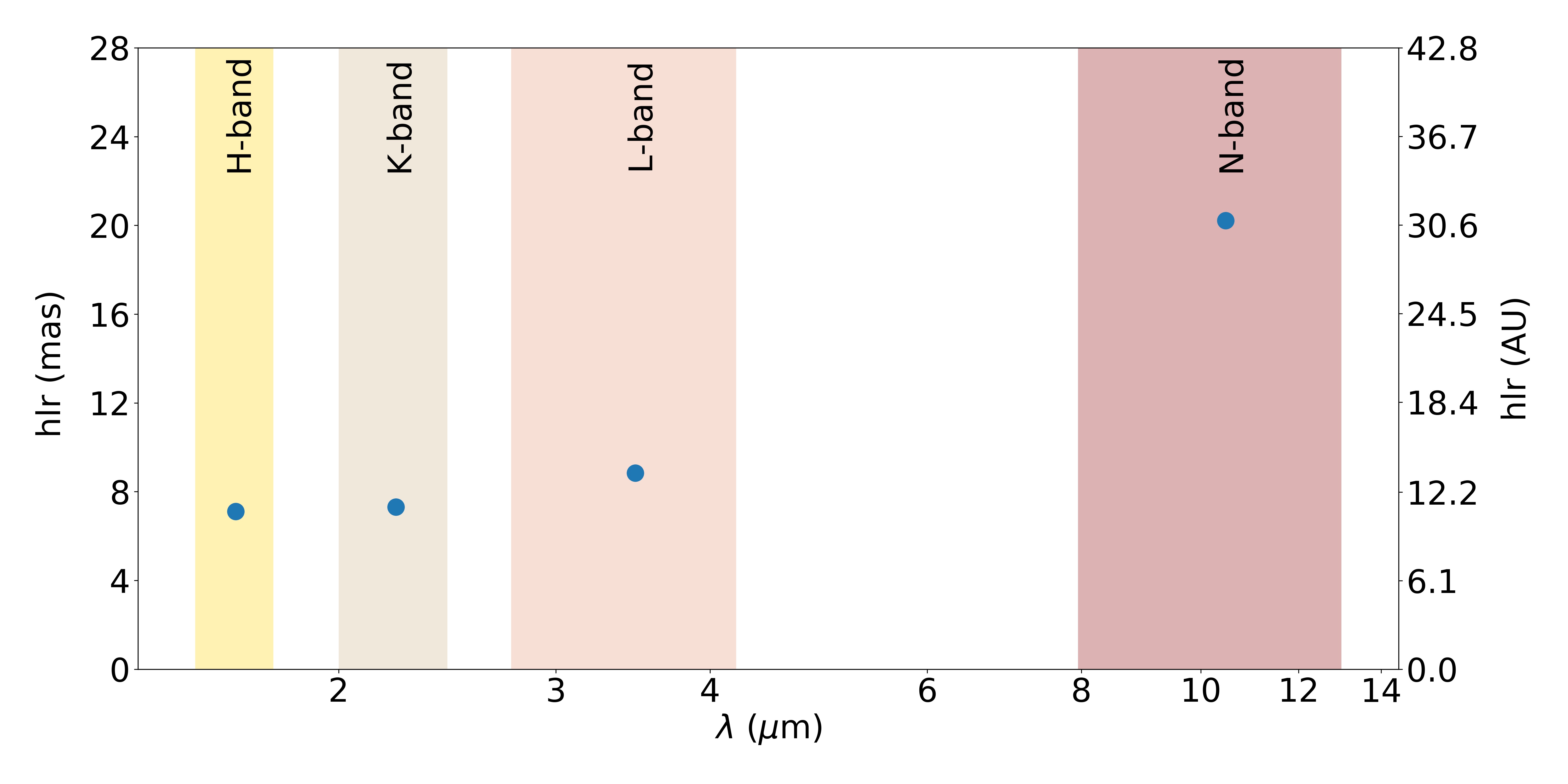}}
  
  \caption{Half-light-radii of the radial emission as a function of wavelength. The half-light radii were calculated at the central wavelengths of each band \modif{for the best-fitting model of each band.}} 
  \label{hlr}
\end{figure}

From the images of the best-fitting models, it is clear that the emission extends farther with increasing wavelength, thus probing a continuous (not necessarily uniform) disc rather than a shell followed by some rings of gas and dust. We can compare the apparent sizes at each band using the half-light-radius (hlr), as shown in Fig. \ref{hlr}. The hlr increases with $
\sim24\%$ from 1.65 $\mu$m to 3.5 $\mu$m and with $\sim 129\%$ between the $L$- and $N$-band.

\section{Discussion}
\label{Sec5}

\modif{The SED of IRAS\,08544 shows that $\sim$ 30\% of the energy emitted by the post-AGB star is captured by the disc \citep{Kluska_2018}. This indicates that the disc is significantly vertically extended at a certain radius. For young stellar objects (YSOs), it is generally accepted that the disc inner rim of the protoplanetary disc (PPD) provides the vertical extension and this disc inner rim is not sharp, but rather curved \citep[e.g.][]{Isella_2005}. For IRAS\,08544, we have a rich data set both in terms of uv-coverage and wavelength range and we show that for the $H$- and $K$-bands, the Gaussian ring model fits the data better, while the flat disc model yields better fits for the $L$- and $N$-bands. The geometry of the Gaussian ring models is more extended as would be expected for rounded inner rims. With our interferometric data set, we interpret that in IRAS\,08544, the vertical structure is located at the disc inner rim and has a similar shape as the ones of PPDs to account for a larger emitting surface at the inner rim (see also Fig. \ref{cut_center_Fig1}). A similar conclusion was reached by \citet{Hillen_2014}, who show that a rounded puffed-up inner rim fits the data of a similar system better, that is the post-AGB binary system 89 Herculis.}

The models in the mid-IR indicate that there is a continuous disc structure in the radial direction. In general, the models fit the overall trend of the visibility data, but they do not fit all data points. This confirms that the disc structure is physically more complex than the geometric models \citep[e.g.][]{Hillen_2017, Kluska_2018}, not only at the inner rim, but also beyond the inner rim. The disc likely has a perturbed asymmetric structure, as was constrained in the near-IR by \citet{Kluska_2019}. Moreover, part of the discrepancy between the geometric models and the long baseline visibility data comes from the accretion disc around the secondary affecting these baselines, while we did not model those effects.

We note that $T_\mathrm{ring}$ is 18\% larger in the $H$-band compared to the $K$-band. In both $H$ and $K$, we probe the hot inner rim, but in $K$ we are also sensitive to parts of the disc slightly beyond the inner rim, where the disc is expected to be colder as the disc temperature decreases as a function of $r$ as in Eq. \ref{Eq: I_nu}. 

\modif{The temperature profile for the $N$-band} is steeper than for the near-IR bands and it is steeper than 0.75 (the maximum expected for protoplanetary discs). \modif{In $N$ we are likely more sensitive to the radial opacity profile in the disc, beyond the inner rim. The study of the full wavelength-dependent opacity profile needs future treatment in radiative transfer models.} 

Circumbinary discs around post-AGB binary systems show many similarities with \modif{PPDs around YSOs} \citep[e.g.][]{deRuyter_2006, Deroo_2006, Deroo_2007, Kluska_2018}. Recently, \citet{Perraut_2019} presented a survey of Herbig Ae/Be stars and concluded that presumably gapped, flared class I sources in the sense of \citet{Meeus_2001} have a large N-to-K band size ratio as opposed to continuous, flat class II sources where the transition is at a value of 10. The N-to-K ratio\modif{, the ratio between the hlr of both bands,} of IRAS\,08544, is \modif{2.8}. Therefore, it might be analogues to a class II source. However, since the transition \modif{may be related to the evolution of the disc \citep{Perraut_2019}, } and as the evolution of circumbinary discs around post-AGB binary systems is not well understood, one needs to check whether this size ratio can also be used for discs around evolving systems in the same way as for protoplanetary discs.

\section{Conclusion} 
\label{Sec6}
Optical interferometric techniques are powerful tools to constrain the compact infrared emission from discs around evolved binary systems. We extended the PIONIER study of IRAS\,08544 to the $K$-, $L$-, and $N$-bands in order to constrain the different flux contributions at different locations in the disc with geometric models. These models fit the overall shape of the visibility data of all VLTI instruments well. The $H$- and $K$-band data were best fitted with a Gaussian ring model, while the $L$- and $N$-band favoured a flat disc model assuming a temperature gradient. \modif{We interpret this as a signature of a puffed-up and rounded inner rim}. In the mid-IR, disc structures beyond the inner rim can be probed and opacity effects start to play an important role. In order to better understand these discs, we not only need to model the radial \modif{extent} of the disc, but also its vertical and azimuthal structure by modelling both visibilities and phase information. This implies using state-of-the-art 3D radiative transfer models to apprehend the full complexity of the disc. The origin of the over-resolved component may be constrained with dedicated disc models \modif{and may shed light on} disc-binary interactions. \modif{Such a rich data set opens up an unprecedented window to precisely study the disc structure and opacity variations at sub-au scales.}

\begin{acknowledgements} 
\modif{We thank the two anonymous referees for their constructive comments and questions that substantially improved the clarity of the paper.} A.C. and H.W.V acknowledge support from FWO under contract G097619N. J.K. acknowledges support from FWO under the senior postdoctoral fellowship (1281121N). This research has benefited from the help of SUV, the VLTI user support service of the Jean-Marie Mariotti Center \footnote{http://www.jmmc.fr/suv.htm}, with special thanks to A. Matter. This research has made use of the Jean-Marie Mariotti Center service \texttt{Aspro} \footnote{Available at http://www.jmmc.fr/aspro} and \texttt{SearchCal} \footnote{Available at http://www.jmmc.fr/searchcal}.
\end{acknowledgements}

%
 


%
%

\begingroup
\bibliographystyle{aa} 
\bibliography{biblio} 

\let\clearpage\relax

\begin{appendix}
\section{Logs of the observations}

\begin{table*}
\caption{Log of GRAVITY observations}\label{GRAVITY_obs}
\centering
\begin{threeparttable}
\begin{tabular}{lcccc}
\hline \hline
Date & Progam ID & MJD & Configuration & Calibrator \\
\hline
2018-11-26(A) & 0102.D-0760 & 58448.31 & A0-G1-J2-J3 & HD\,75063\\
2018-11-26(B) & 0102.D-0760 & 58448.34& A0-G1-J2-J3 & HD\,75789\\
2018-12-23 & 0102.D-0760 &58475.18& A0-B2-C1-D0 & HD\,75063 \\
2019-01-14 & 0102.D-0760 & 58497.35 & D0-G2-J3-K0 & HD\,79940\\ 

\hline
\end{tabular}
\end{threeparttable}
\end{table*}

\begin{table*}
\caption{Log of MATISSE observations}\label{MATISSE_obs}
\centering
\begin{threeparttable}
\begin{tabular}{lccccc}
\hline \hline
Date & Progam ID & MJD & Configuration & Calibrator(s) L & Calibrator(s) N\\
\hline
2019-04-29(A) &60.A-9275 &58602.06& D0-G2-J3-K0 & IRAS\,10153-5540 & IRAS\,10153-5540\\
2019-04-29(B) &60.A-9275 &58602.08& D0-G2-J3-K0 & IRAS\,10153-5540 & IRAS10153-5540\\
2019-05-01 &60.A-9275 &58604.02&D0-G2-J3-K0 & HD\,86355 + IRAS\,08534-2405 & IRAS\,08534-2405\\
2019-05-02&60.A-9275 & 58605.08& A0-G1-J2-J3 & HD\,74600 & HD\,74600\\
2019-05-04&60.A-9275 &58606.99 & A0-G1-J2-J3 & IRAS\,10153-5540 + IRAS\,08534-2405& IRAS\,10153-5540 \\
2019-05-09 &60.A-9275 & 58612.95& A0-B2-C1-D0 & IRAS\,08534-2405 & IRAS\,08534-2405\\
2019-05-11(A) &60.A-9275 & 58614.07 & A0-B2-C1-D0 & HIP\,55282 + HD\,123139 & HIP\,55282\\
2019-05-11(B) &60.A-9275 & 58614.08 & A0-B2-C1-D0 & HIP\,55282 + HD\,123139 & HD\,123139\\
2020-01-21& 0104.D-0739 &  58869.08 & A0-G1-J2-K0& HD\,74600 & HD\,74600 \\
2020-03-01(A)& 0104.D-0739 & 58909.02 & A0-G1-J2-J3 & HD\,86355 + CD-55\,3254 & HD\,86355 + CD-55\,3254\\
2020-03-01(B) & 0104.D-0739 & 58909.14& A0-G1-J2-J3& HD\,86355 + CD-55\,3254 & HD\,86355 + CD-55\,3254\\

\hline
\end{tabular}
 \begin{tablenotes}
 \item 
    \end{tablenotes}

\end{threeparttable}
\end{table*}

\begin{table*}

\caption{Information of the calibrators for the N-band}\label{N-band_cal_info}

\begin{threeparttable}
\begin{tabular}{lcccc}
\hline \hline 
Calibrator & distance (pc) \tablefootmark{a}  & Diameter (mas) & Spectral type & $T\mathrm{eff}$ (K) \textbf{\tablefootmark{b}}\\
\hline
IRAS\,10153-5540 & 2302.0$^{+105.3}_{-125.4}$ (1)& 2.748 $\pm$ 0.270 $(2)$ &M3 $(3)$ & \modif{3400}\\
HD\,74600  & 1069.3$^{+65.3}_{-61.0}$ (1) &2.940 $\pm$ 0.259 (2) & M3Ib/II (4) & \modif{3400} \\
HD\,123139 & 18.0 $\pm$ 0.07 (5) &5.517 $\pm$ 0.434 $(2)$&  KOIII (6) & \modif{5000}\\ 
IRAS\,08534-2405 & 886.5 $^{+54.2}_{-57.4}$ (1)& 3.613 $\pm$ 0.254 \tablefootmark{c} & M7 (7) & \modif{3100}\\
HIP\,55282 & 58.62  $^{+0.52}_{-0.50}$ (1) & 3.314 $\pm$ 0.356 (2) & G9 III (5) & \modif{5200}\\
\hline
\end{tabular}
\tablebib{ (1) \citet{Bailer-Jones_2021};
(2) \citet{bourges_2017};
(3) \citet{Blanco_1955};
(4) \citet{Houk_1978};
(5) \citet{vanLeeuwen_2007};
(6) \citet{Keenan_1989};
(7) \citet{Hansen_1975}}
\tablefoot{ \tablefootmark{a}. The entry 'geometric distance' in the distance catalogue of \citet{Bailer-Jones_2021} \modiff{was used}, except for HD\,123139 which is not in Gaia EDR3 \modif{and therefore not \modiff{processed} by \citet{Bailer-Jones_2021}. In that case, we used the Hipparcos parallax to estimate the distance}. The uncertainties are 1$\sigma$. 
\\ \tablefootmark{b} \modif{The effective temperatures of the calibrators} were estimated from the spectral type using \citet{Pickels_1998}.
\\ \tablefootmark{c} IRAS\,08534-2405 is not in the JMMC Stellar Diameter Catalog (JSDC). The diameter was taken from getStar (https://apps.jmmc.fr/~sclws/getstar/).
}

\end{threeparttable}
\end{table*}

\section{Additional table}
\begin{table*}

\caption{Accepted ranges of the different fitting parameters}
\label{ranges of parameters}

\begin{tabular}{ lcc }
\hline \hline

Parameter & Range \\
\hline

f$_{star_0} (\%)$ & 0 - 100 \\
f$_{back_0}$ (\%) & 0 - 100 \\ 
$d_\mathrm{bg}$ & $10 - 0$\\
$\theta$ (mas) & $10 - 40$ \\
$w$ (mas) & $3 - 25$\\
$T_\mathrm{ring}$ (K) & $200 - 2000 $\\
$T_\mathrm{in}$ (K)& $1000 - 2000$\\
$q$ & $10 - 0$ \\
$R_\mathrm{in}$ (mas)& $0 - 25$\\
$F_\mathrm{tot}$ (Jy)&  $ 100 - 400$\\
\hline
\end{tabular}
\end{table*}

\section{Additional figures}
\begin{figure*}[htb]
\centering
    \begin{minipage}[t]{0.49\textwidth}
 \includegraphics[width=\textwidth,width=1.0
  \textwidth]{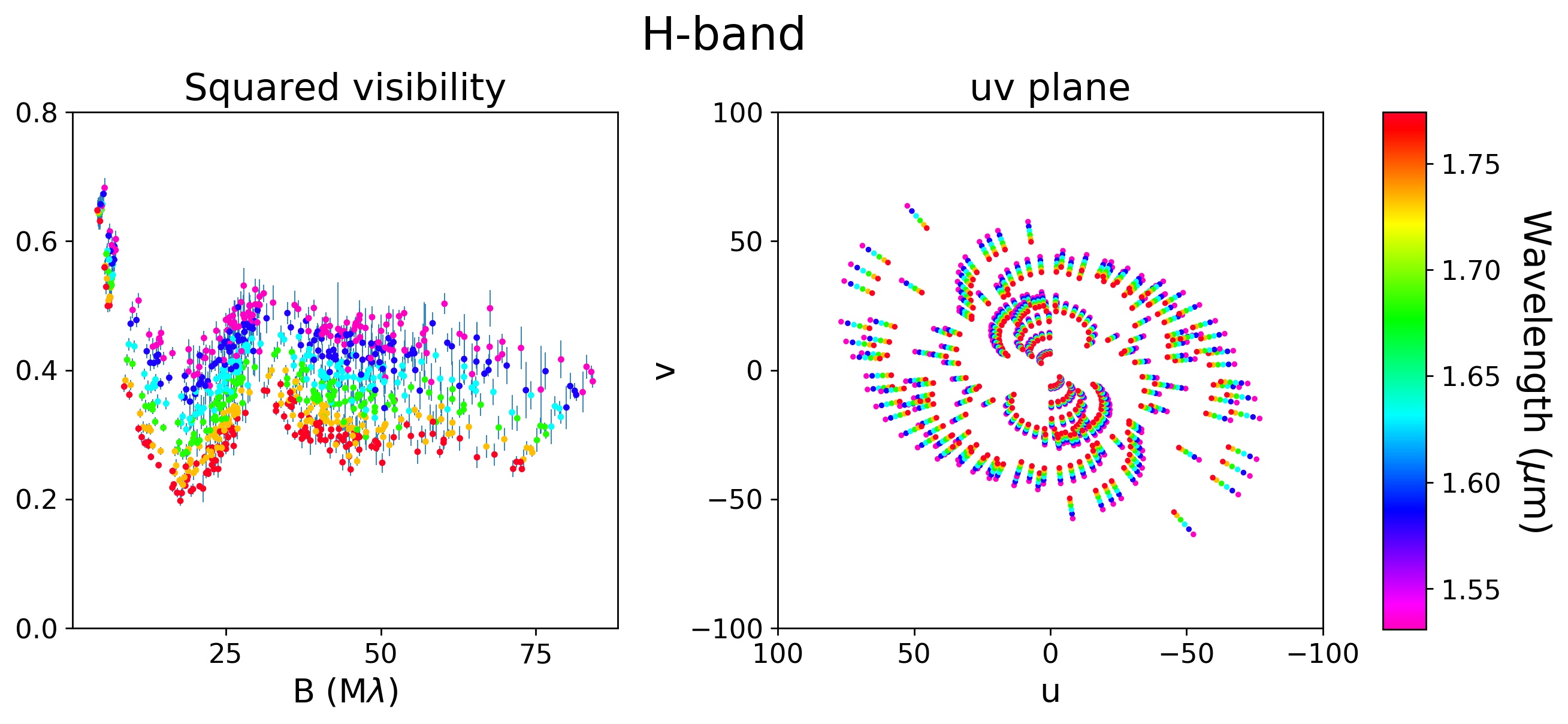}
  \end{minipage}%
  \hfill
  \begin{minipage}[t]{0.49\textwidth}
 \includegraphics[width=\textwidth,width=01.0
  \textwidth]{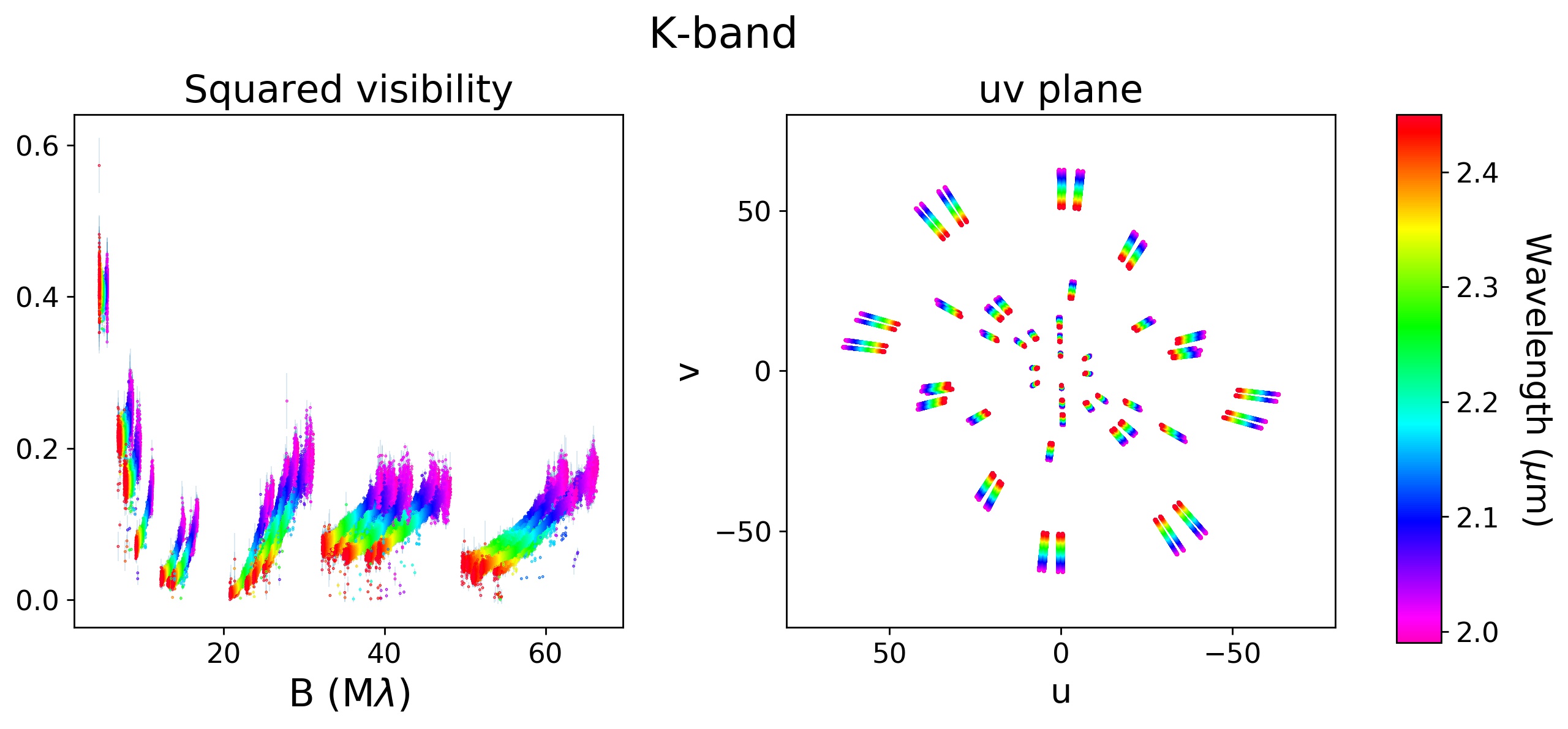}
  \end{minipage}
  
  \begin{minipage}[b]{0.49\textwidth}
 \includegraphics[width=\textwidth,width=1.0
  \textwidth]{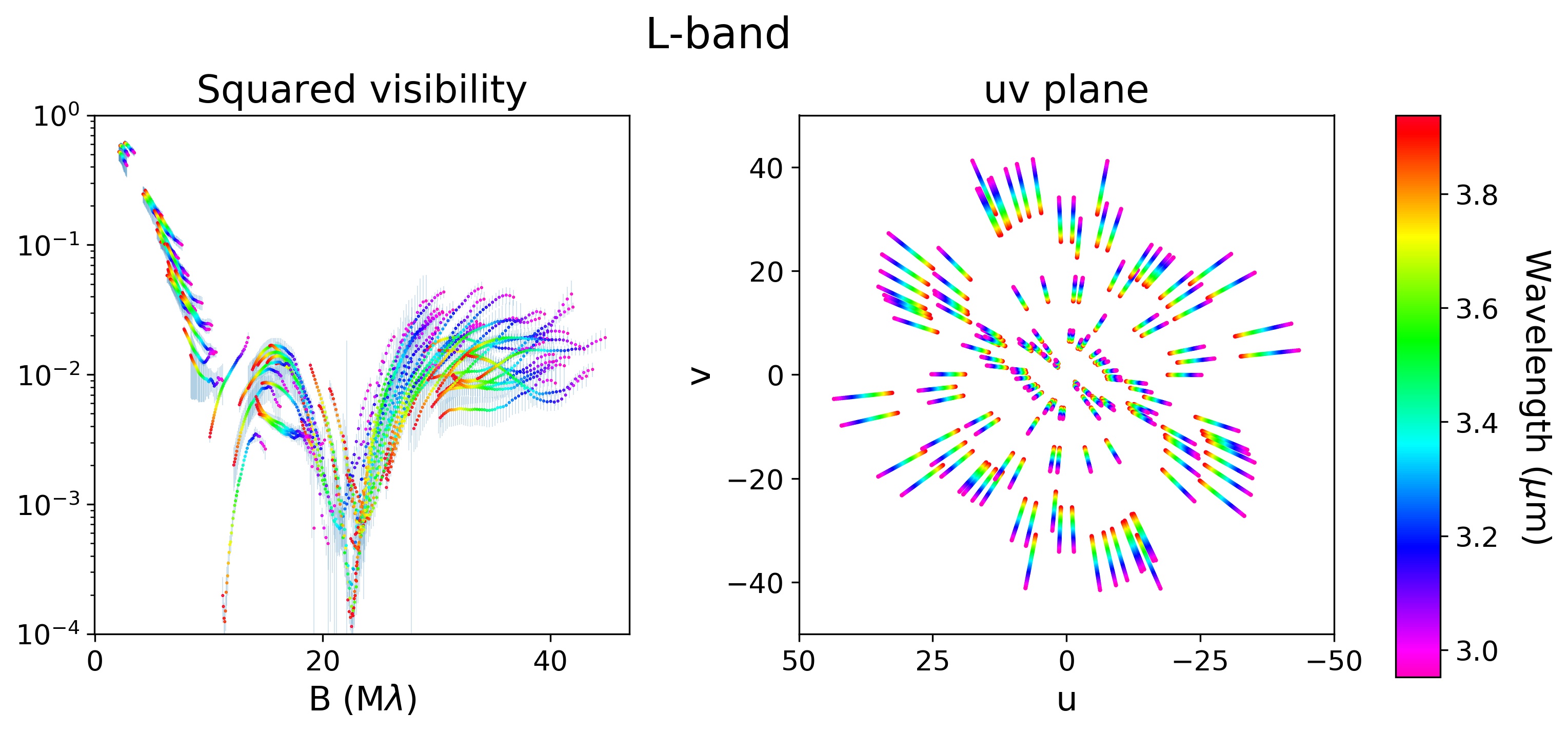}
  
  \end{minipage}%
  \hfill
  \begin{minipage}[b]{0.49\textwidth}
 \includegraphics[width=\textwidth,width=1.0
  \textwidth]{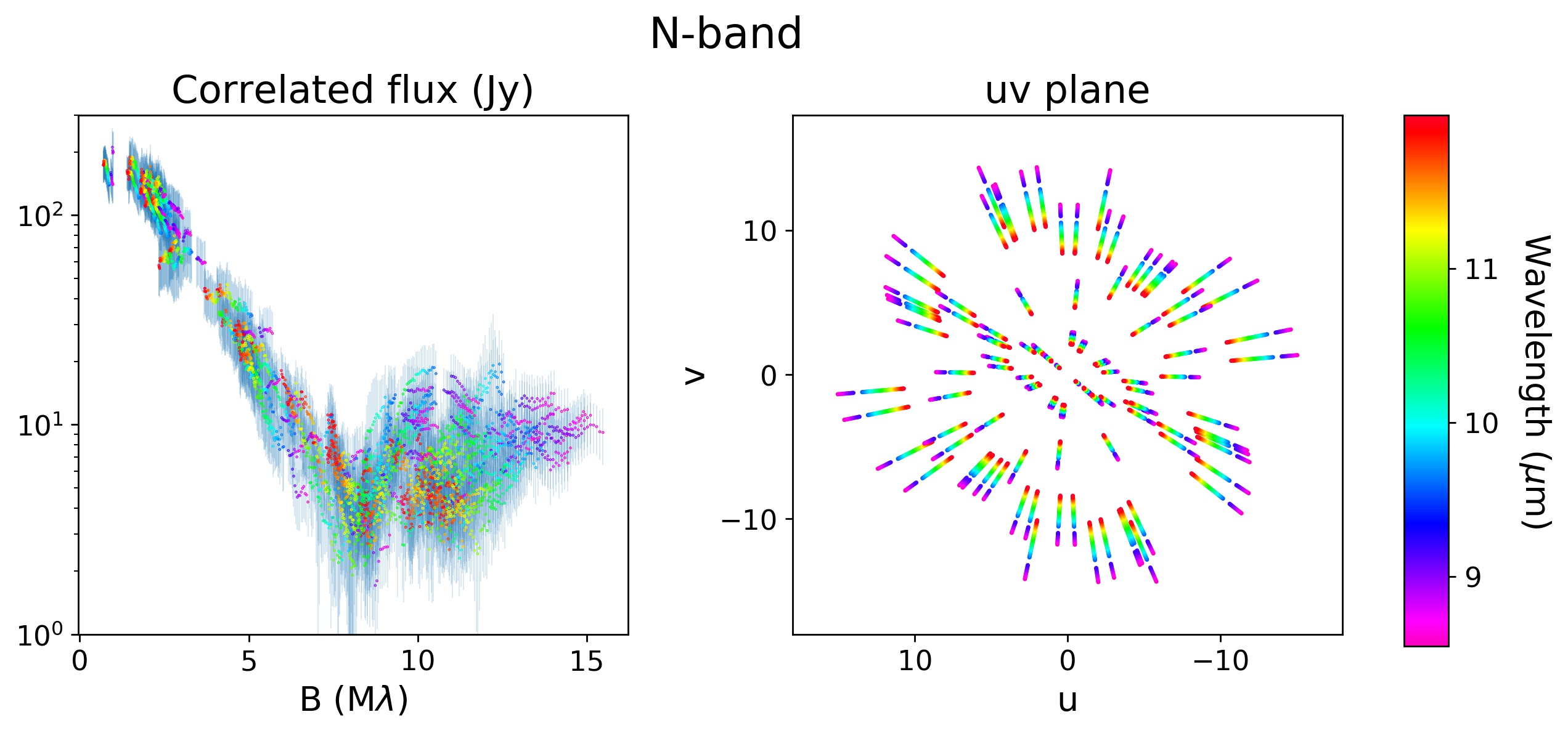}
  \end{minipage}
  \caption{Visibility data as a function of the baseline and wavelength and the uv-coverages of the accepted data. Top left: $H$-band. Top right: $K$-band. Bottom left: $L$-band. Bottom right: $N$-band. The near-IR data \modif{show} a strong chromatic effect in the squared visibilities, whereas there is a mono-chromatic effect at wavelengths \modif{at which} the circumbinary disc dominates.}
   \label{uv_coverages}
\end{figure*}

\begin{figure*}

\centering
\begin{minipage}{0.49\textwidth}
  \includegraphics[width=\textwidth,width=1.0
  \textwidth]{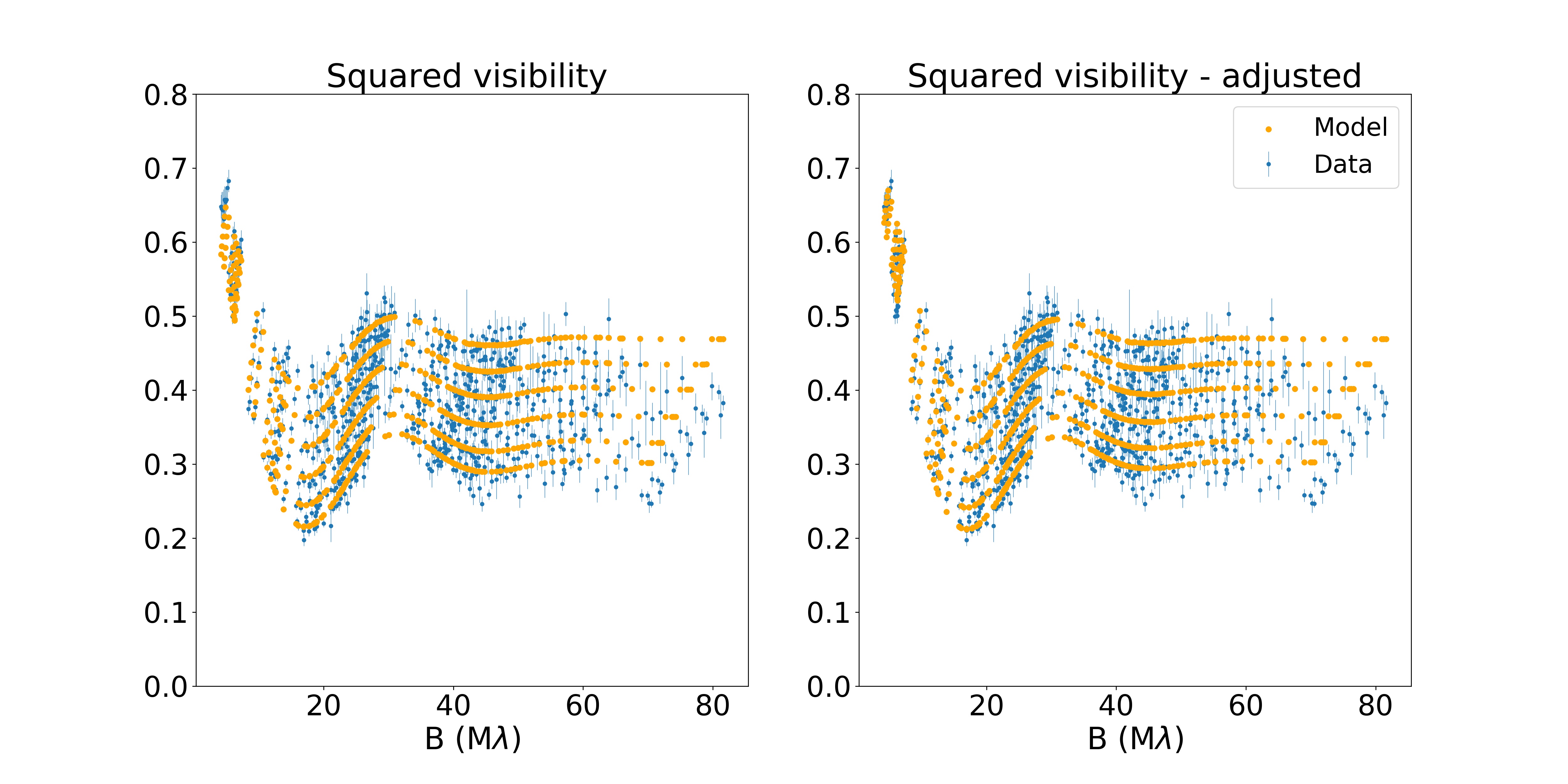}
  \end{minipage}
  \begin{minipage}{0.49\textwidth}
  \includegraphics[width=\textwidth,width=1.0
  \textwidth]{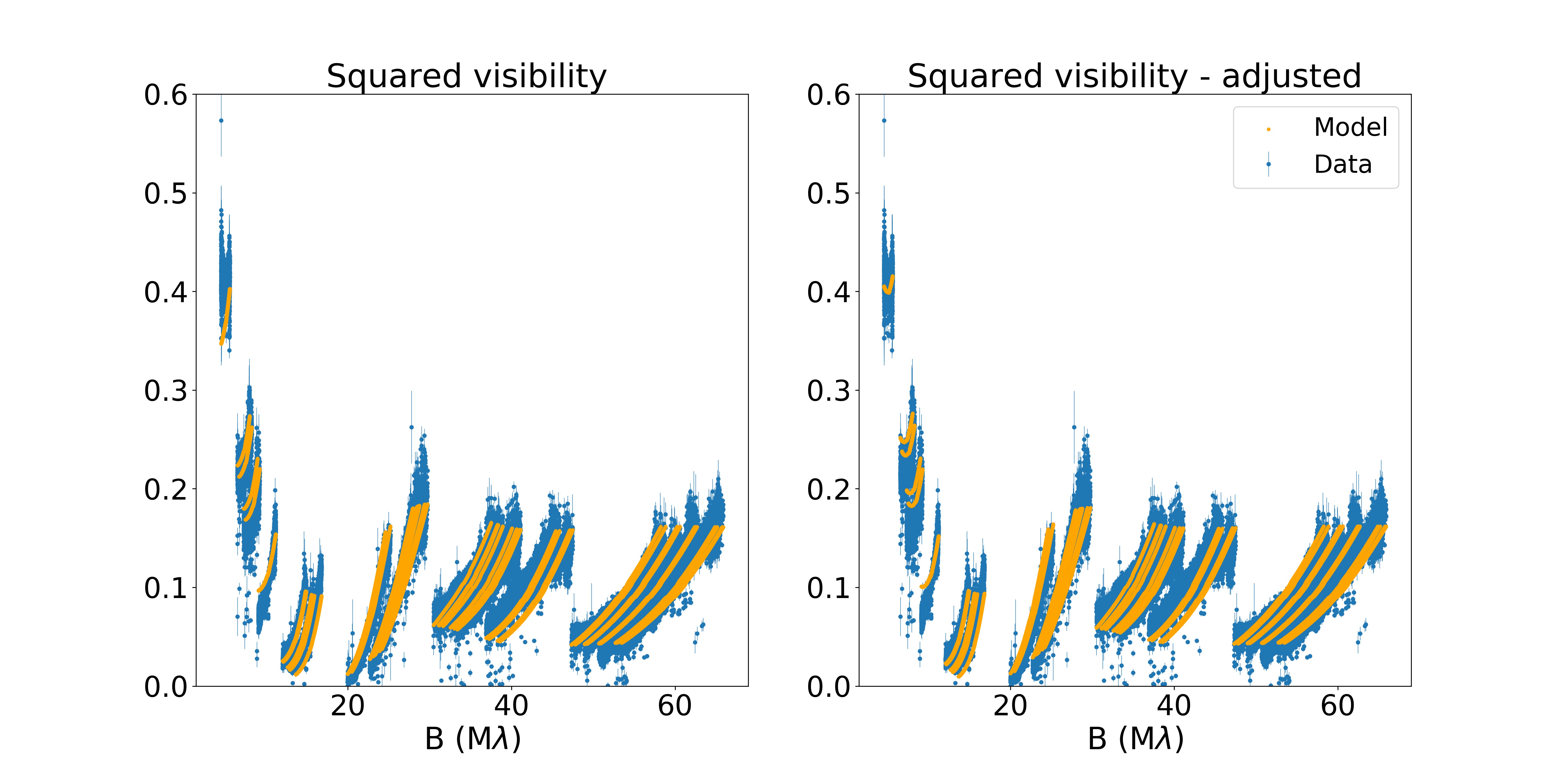}
  \end{minipage}
\begin{minipage}{0.49\textwidth}
  \includegraphics[width=\textwidth,width=1.0
  \textwidth]{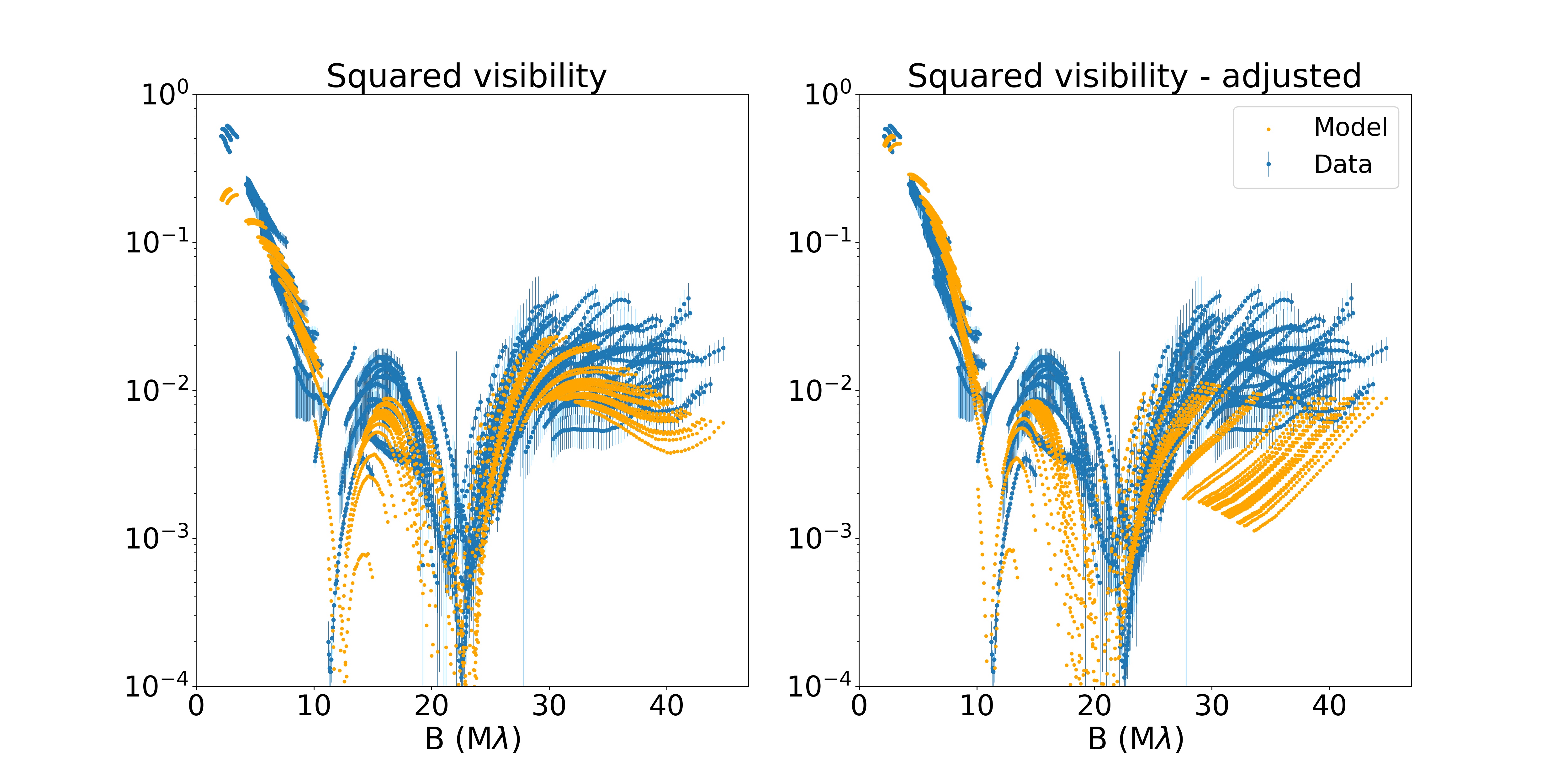}
  \end{minipage}
  \begin{minipage}{0.49\textwidth}
  \includegraphics[width=\textwidth,width=1.0
  \textwidth]{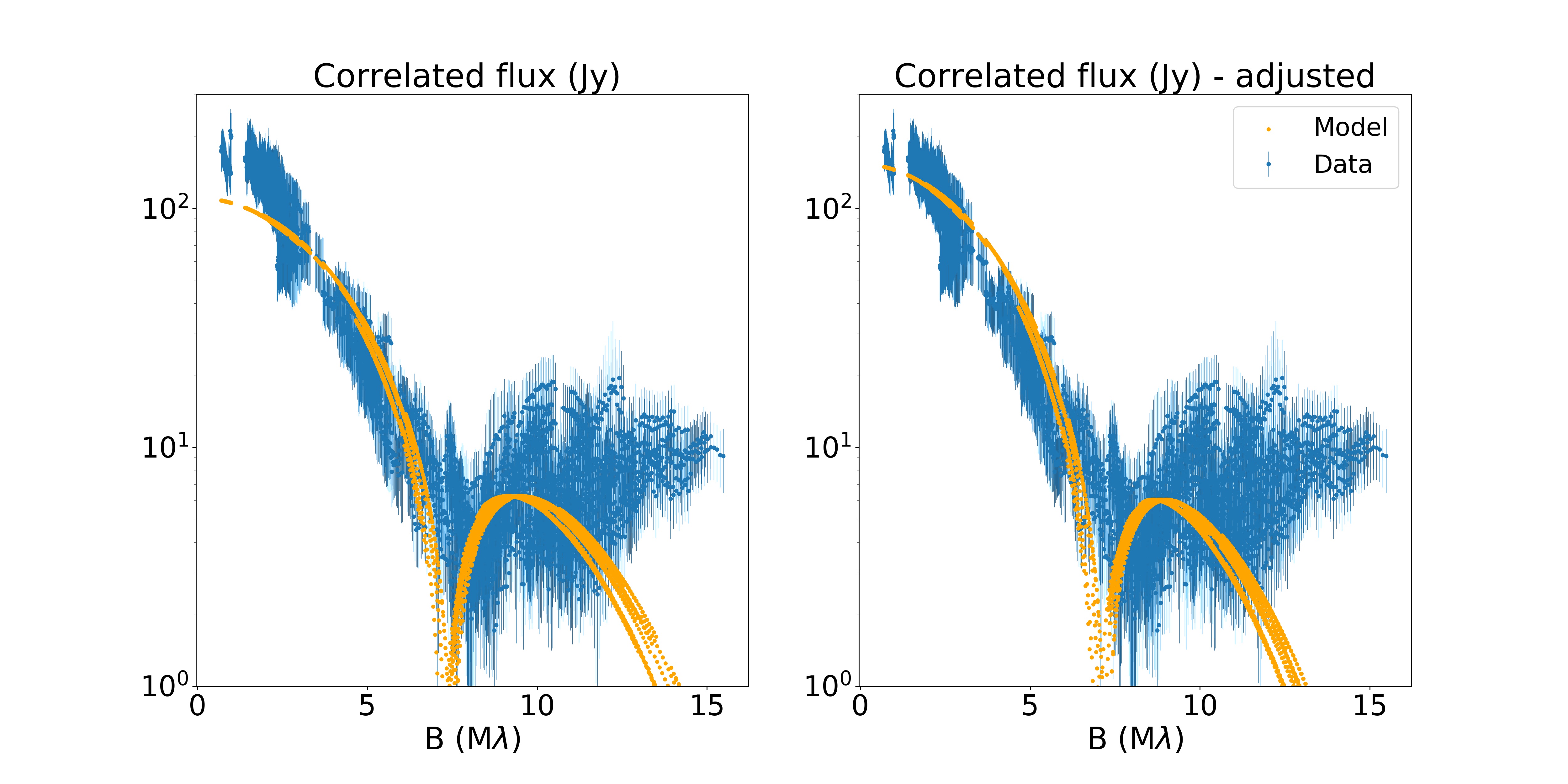}
  \end{minipage}
  
  \caption{Illustration of the effect of increasing the weight of the shortest baselines in order to fit those baselines. Top left: $H$-band. Top right: $K$-band. Bottom left: $L$-band. Bottom right: $N$-band. The left panels show the fit of the Gaussian ring model while leaving the data as they are. The right panels show the fit when the weight of the shortest piece of baseline data is increased. Only the effect on the Gaussian ring model is shown. For $H$ ($B< 8$ M$\lambda$), $K$ ($B< 5.5$ M$\lambda$), and $L$ ($B< 3.5$ M$\lambda$), the effect of increasing the weight of the shortest baselines is the same for both the Gaussian ring model and the flat disc model, while the shortest baselines were fitted in the flat disc model without a modification on the weights of the shortest baselines ($B< 2.5$ M$\lambda$) for the $N$-band. By increasing the weight on the shortest baselines for $L$, the shape of the longer baselines was shifted such that overall lower visibilities are predicted. This latter effect does not arise when fitting with the flat disc model.}
  \label{short_baselines_comp}
\end{figure*}

\begin{figure*}

\centering

  \includegraphics[width=\textwidth,width=0.9
  \textwidth]{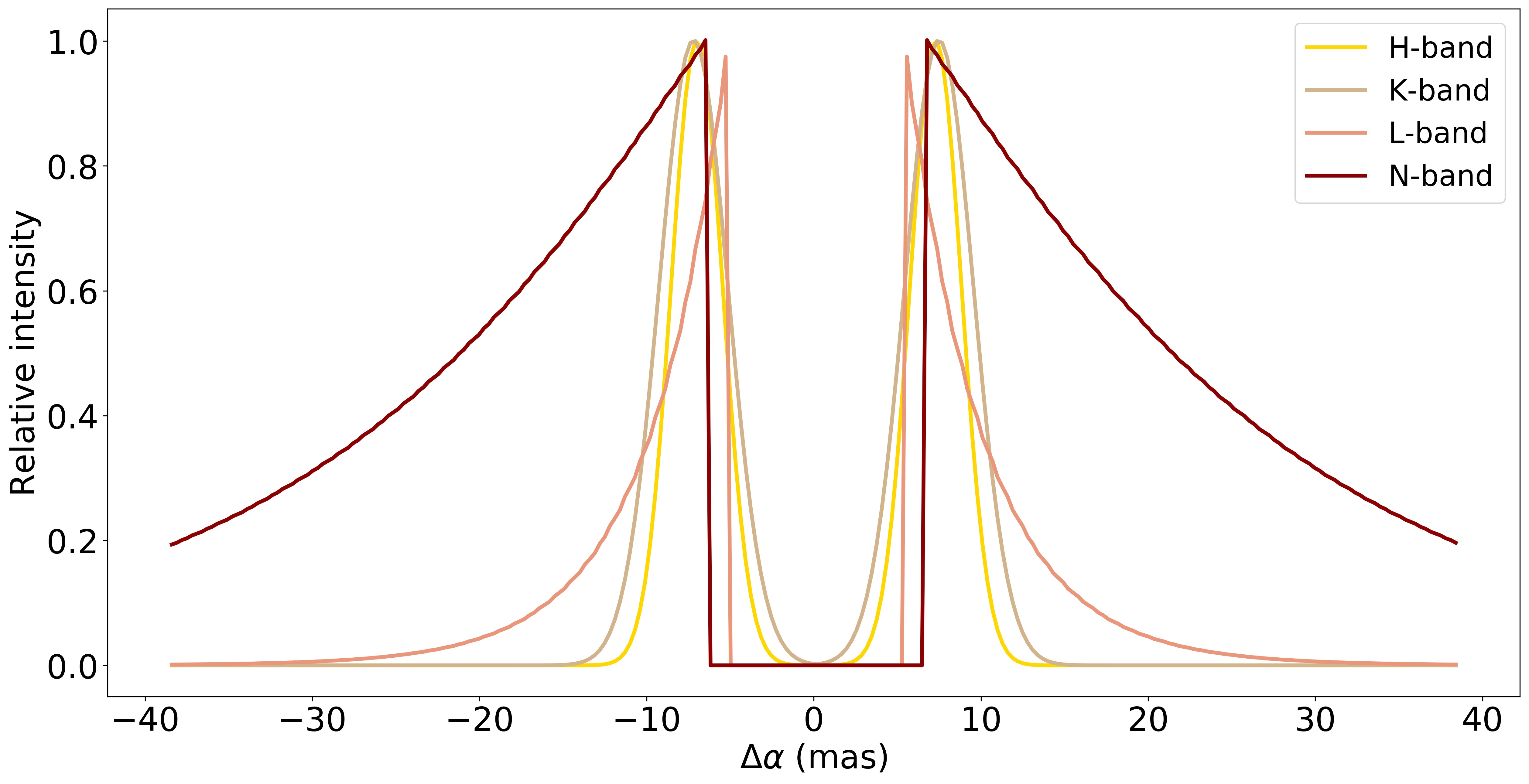}
 
  \caption{\modif{Radial cuts through the centre of the images of the best-fitting geometric models of Fig. \ref{fig:Images_all_bands} showing the profiles, the inner holes, and the extent of the emission of the individual bands. The best-fitting model for the $H$- and $K$-bands is a Gaussian ring model. The best-fitting model of the $L$- and $N$-bands is the flat disc model which has an inner hole of $R_\mathrm{in}$.}}
  \label{cut_center_Fig1}
\end{figure*}
\end{appendix}
\endgroup

\end{document}